\documentclass[aps,twocolumn, superscriptaddress]{revtex4-2}

\usepackage[hidelinks]{hyperref} 
\usepackage{graphicx}
\usepackage{amsmath}
\usepackage{amssymb}
\usepackage{siunitx}
\usepackage{braket}
\usepackage{bm}
\usepackage{mathtools}

\DeclareGraphicsExtensions{.png,.jpg,.eps}
\usepackage{xcolor}
\hypersetup{
    colorlinks,
    linkcolor={red!50!black},
    citecolor={blue!80!black},
    urlcolor={blue!80!black}
}

\def\m1{{^{-1}}}

\begin{document}

\author{Eric G. Schertenleib}
\affiliation{Institute for Theoretical Physics, ETH Zurich, 8093 Zurich, Switzerland}

\author{Mark H. Fischer}
\affiliation{Department of Physics, University of Zurich, 8057 Zurich, Switzerland}

\author{Manfred Sigrist}
\affiliation{Institute for Theoretical Physics, ETH Zurich, 8093 Zurich, Switzerland}

\title{
Unusual $H$-$T$ phase diagram of CeRh$_2$As$_2$ - the role of staggered non-centrosymmetricity  
}

\begin{abstract}
Superconductivity in a crystalline lattice without inversion is subject to complex spin-orbit-coupling effects, which can lead to mixed-parity pairing and an unusual magnetic response. In this study, the properties of a layered superconductor with alternating Rashba spin-orbit coupling in the stacking of layers, hence (globally) possessing a center of inversion, is analyzed in an applied magnetic field, using a generalized Ginzburg-Landau model. The superconducting order parameter consists of an even- and an odd-parity pairing component which exchange their roles as dominant pairing channel upon increasing the magnetic field. This leads to an unusual kink feature in the upper critical field and a first-order phase transition within the mixed phase. We investigate various signatures of this internal phase transition. The physics we discuss here could explain the recently found $H$--$T$ phase diagram of the heavy Fermion superconductor CeRh$_2$As$_2$.
\end{abstract}

\date{\today}

\maketitle


\section{Introduction}

The discovery of the heavy Fermion superconductor CePt$_3$Si~\cite{PhysRevLett.92.027003} triggered much interest in superconductivity in materials lacking an inversion center in their crystalline structure. Among the most intriguing features of such non-centrosymmetric superconductors we may mention the parity mixing of Cooper-pair states~\cite{JETP.68.1244, PhysRevLett.87.037004,sigristbook}, the helical superconducting phase induced by a magnetic field~\cite{PhysRevLett.92.027003, JETP.78.401, PhysRevLett.94.137002}, and the appearance of topological superconducting phases~\cite{PhysRevB.79.094504, PhysRevLett.103.020401}. These properties emerge most spectacularly, if the superconductor is intrinsically prone to unconventional pairing as expected for heavy-Fermion compounds due to strong electron correlation. 
 
Parity mixing is generated by anti-symmetric spin-orbit coupling present in non-centrosymmetric crystals, with Rashba spin-orbit coupling in systems which lack mirror symmetry the best-known example. This type of spin-orbit coupling is realized in a number of Ce-based heavy Fermion superconductors with tetragonal crystal symmetry, such as CeIrSi$_3$~\cite{JPSJ.76.044708} and CeRhSi$_3$~\cite{PhysRevLett.95.247004}, as well as CePt$_3$Si. 
The probably most striking properties seen in both, CeIrSi$_3$ and CeRhSi$_3$ is the enormous upper critical field for field directions perpendicular to the basal plane (lacking mirror symmetry), exceeding the paramagnetic limit by far ~\cite{JPSJ.77.073705,PhysRevLett.98.197001}. This indicates also an extremely short coherence length which is not unusual for heavy-Fermion superconductors. 
The isostructural La-versions of these two compounds (replacing Ce by La), which superconduct with comparable critical temperature but without heavy-Fermion properties, have low upper critical fields dominated by orbital depairing~\cite{sigristbook}. The interplay of orbital and paramagnetic depairing has played an important role in other Ce-based superconductors, too, most notably in the centrosymmetric CeCoIn$_5$, which besides displaying paramagnetic limiting behavior is also well-known for its low-temperature high-magnetic field phase, the so-called Q-phase, where superconductivity coexists with a spin density wave state~\cite{PhysRevLett.91.187004,Kenzelmann1652}. This indicates that the Maki parameter $\alpha_M = \sqrt{2} H_{c2}(0)/H_p(0)$ is not small in this material, where $H_{c2}$ and $H_p$ denote the paramagnetic and orbital upper critical field, respectively,~\cite{PhysRev.148.362}. 
 
Recently, it has been noticed that centrosymmetric materials may incorporate locally non-centrosymmetric units which can influence properties of a superconductor profoundly, in particular their behavior in a magnetic field~\cite{PhysRevB.84.184533,JPSJ.83.061014}. Features of this origin are found, for example, in superlattices such as the regular stacks of superconducting CeCoIn$_5$ alternating with layers of YbCoIn$_5$~\cite{Science.327.980, Nature.7.849}.
Here, local non-centrosymmetricity at the interfaces between Ce- and Yb-layers involves antisymmetric spin-orbit coupling which apparently protects Cooper pairs against paramagnetic depairing in this high-$\alpha_M $ system~\cite{JPSJ.81.034702}. Moreover, it was suggested that a peculiar parity-mixing present in these superconductors~\cite{PhysRevB.84.184533} may give rise to a magnetic field phase transition changing the character of the pairing state~\cite{PhysRevB.86.134514}. 

The very recently discovered heavy-Fermion superconductor CeRh$_2$As$_2$ belongs also to the class of locally non-centrosymmetric superconductors. While its tetragonal crystal structure has an inversion center, it consists of layers with alternating violation of inversion symmetry. The upper critical field directed along the $c$-axis (perpendicular to the staggered layers) extrapolates to $ \sim 10\, \mathrm{T}$ at zero temperatures, which lies far beyond the paramagnetic limiting field $ H_p \sim 0.5 \, \mathrm{T}$ for a critical temperature $ T_c \lesssim 0.4 \mathrm{k} $~\cite{hassinger}. The most intriguing aspect of this superconductor so far is its behavior in the $c$-axis magnetic field. The upper critical field shows a pronounced kink for a temperature $ T$ roughly half of $ T_c $. This anomaly strongly hints towards a switch in the order parameter symmetry upon increasing magnetic field. 

Stimulated by this experimental finding, we investigate here a possible scenario explaining this behavior based on a model of a material with staggered Rashba spin-orbit coupling~\cite{PhysRevB.84.184533}. This symmetry property can be easily implemented into a generalized Ginzburg-Landau theory which is well suited to discuss the superconducting order parameter in the mixed phase. The basic idea of a field-induced phase transitions has been previously explored by Yanase and co-workers in the context of the Ce/YbCoIn$_5$ superlattices using a Bogolyubov-de Gennes approach which ignored the presence of vortices~\cite{JPSJ.81.034702,JPSJ.82.074714,PhysRevB.86.134514}. In these studies, a few superconducting layers were considered whose 
 two boundary layers constitute opposite non-centrosymmetric environments and, thus, Rashba spin-orbit coupling of opposite sign. The subsequent extension to the mixed phase was performed by M\"ockli \textit{et al.} using a Ginzburg-Landau model~\cite{PhysRevB.97.144508}. In our investigation we base our analysis of the infinite-layer system with staggered Rashba spin-orbit coupling on this approach to analyze the situation of CeRh$_2$As$_2$. 
 
 In the following, we first introduce the Ginzburg-Landau model describing the effect of parity mixing by including an even- and odd-parity order parameter component. We show that this allows us to discuss two phases, which we refer to as A and B phase, which, in view of the crystalline inversion center, can be considered as even or odd, respectively. In a magnetic field along the crystalline $c$ axis, we can reproduce qualitatively the kink feature of the upper critical field and also find an internal first-order phase transition within the mixed phase. The mixed phase is described here in a scheme using a cellular approximation of the vortex lattice unit cell~\cite{PhysRevB.64.064517}. Finally, we propose several ways to detect the internal phase transition in experimental.


\section{Ginzburg-Landau model free energy}

\subsection{Free energy functional and pairing symmetries}
\label{subsec: Free energy functional and pairing symmetries}

Even though the crystal lattice of our system possesses global inversion symmetry and a superconducting order parameter can thus be classified as even or odd under inversion, the unit cell of the superlattice includes two locally non-centrosymmetric  subsystems stacked alternatingly on top of each other. The lack of local inversion symmetry gives rise to a staggered Rashba spin-orbit coupling, leading to the formation of Cooper pairs of mixed parity within each layer~\cite{PhysRevLett.87.037004,PhysRevB.84.184533, JPSJ.83.061014}. Within the Ginzburg-Landau formalism, this is accounted for by introducing even- and odd-parity order-parameter components on each layer $j$, $\Psi_{e,j}(\bm{r}, \varphi_{e,j}) = \psi_{e,j}(\bm{r}) \exp(i\varphi_{e,j})$ and $\Psi_{o,j}(\bm{r},\varphi_{o,j})=\psi_{o,j}(\bm{r}) \exp (i \varphi_{o,j})$, respectively, with $ \bm{r} $ the in-plane coordinate. 

We write the free energy functional as
\begin{equation}
    F[\Psi_e, \Psi_o, \bm{A}] =  \sum\limits_j \int d^2r f^{(j)}(\bm{r}),
\end{equation}
with $\bm{A}$ denoting the vector potential. 
The free energy density of the $j$th layer takes the form,

\begin{eqnarray}
f^{(j)}(\bm{r}) &=&\frac{\bm{B(\bm{r})}^2}{8\pi}+  \sum\limits_{l=e,o}\bigg[-a_l|\Psi_{l,j}(\bm{r})|^2 + \frac{b_l}{2}|\Psi_{l,j}(\bm{r})|^4  \nonumber \\
&+& \frac{1}{2m_l} \left|\bm{D}_\parallel  \Psi_{l,j}(\bm{r})\right|^2 + \tilde{J}_l \left|\Psi_{l,j+1}(\bm{r})-\Psi_{l,j}(\bm{r})\right|^2 \bigg]\nonumber \\
&+&\frac{\tilde{Q}}{2} \bm{B}^2(\bm{r}) |\Psi_{e,j}(\bm{r})|^2 +f_{eo}^{(j)}(\bm{r}) \label{eq: free_energy_density}
\end{eqnarray}
with $\bm{B} = \nabla \times \bm{A}$ the magnetic induction perpendicular to the plane and the summation running over both the even ($l=e$) and odd ($l=o$) pairing channels. The parameters $a_l=a_{0,l}(T_{c,l}-T)$, with $T_{c,l}$ representing the bare critical temperature of the respective pairing channel, and $b_l$ are phenomenological constants. We choose both critical temperature $T_{c,l} \geq 0$, indicating that the corresponding Cooper pairing channels have comparable attractive interactions, but with
$ T_{c,e} >T_{c,o} $. The parameter $m_l$ represents the in-plane  Cooper-pair mass of the respective order-parameter component and the covariant derivative $\bm{D}_\parallel = (-i \hbar \nabla +2e\bm{A}(\bm{r})/c)_\parallel$ is restricted to the two in-plane coordinates. The last term in the sum incorporates the coupling of the order parameters between neighboring layers with coupling strength $\tilde{J}_l$, assuming a quasi-two-dimensional electronic structure.  Furthermore, we included the paramagnetic depairing effect, which directly affects the even-parity component and whose strength is determined by the parameter $\tilde{Q}$~\cite{JPSJ.81.093701, sigristbook, PhysRevB.86.134514}. Note, however, that due to the parity mixing $f_{eo}^{(j)}$ discussed below, paramagnetic depairing is effectively detrimental to both order-parameter components. We assume the field to be equal for all layers.

To lowest order, the spin-orbit-coupling-induced parity mixing between even and odd components in the $j$th layer takes the form~\cite{JPSJ.83.044712}
\begin{eqnarray}
    f_{eo}^{(j)} &=& \frac{\tilde{\epsilon}_j}{2} (\Psi_{e,j}\Psi_{o,j}^* + \Psi_{e,j}^* \Psi_{o,j})
    \label{eq: free_energy_fst_1}\nonumber\\
    &=& \tilde{\epsilon}_j|\Psi_{e,j}||\Psi_{o,j}| \cos \left(\varphi_{e,j}-\varphi_{o,j}\right), 
\end{eqnarray}
with $\tilde{\epsilon}_j$ the coupling strength on the $j$th layer, which, due to the staggered nature of the spin-orbit coupling, possesses alternating sign on neighboring layers, i.e. $\tilde{\epsilon}_j = (-1)^j \tilde{\epsilon}$ with $\tilde{\epsilon}> 0 $. The free energy is constructed in a way to be invariant under both time-reversal symmetry (TRS) $\mathcal{T}$
\begin{subequations}
\begin{align}
\Psi_e,\, \Psi_o &\stackrel{\mathcal{T}}{\longrightarrow} \Psi_e^*, \, \Psi_o^*,\\
\bm{A}  &\stackrel{\mathcal{T}}{\longrightarrow} -\bm{A},
\end{align}
as well as local $\mathrm{U}(1)$ gauge transformations $\Phi$
\begin{align}
\Psi_e, \, \Psi_o &\stackrel{\Phi}{\longrightarrow} e^{i\phi(x)} \Psi_e,\, e^{i\phi(x)} \Psi_o \\
\bm{A} &\stackrel{\Phi}{\longrightarrow} \bm{A} + \nabla \phi(x).
\end{align}
\end{subequations}

\begin{figure}[b]
\begin{center}
\includegraphics[width=\linewidth]{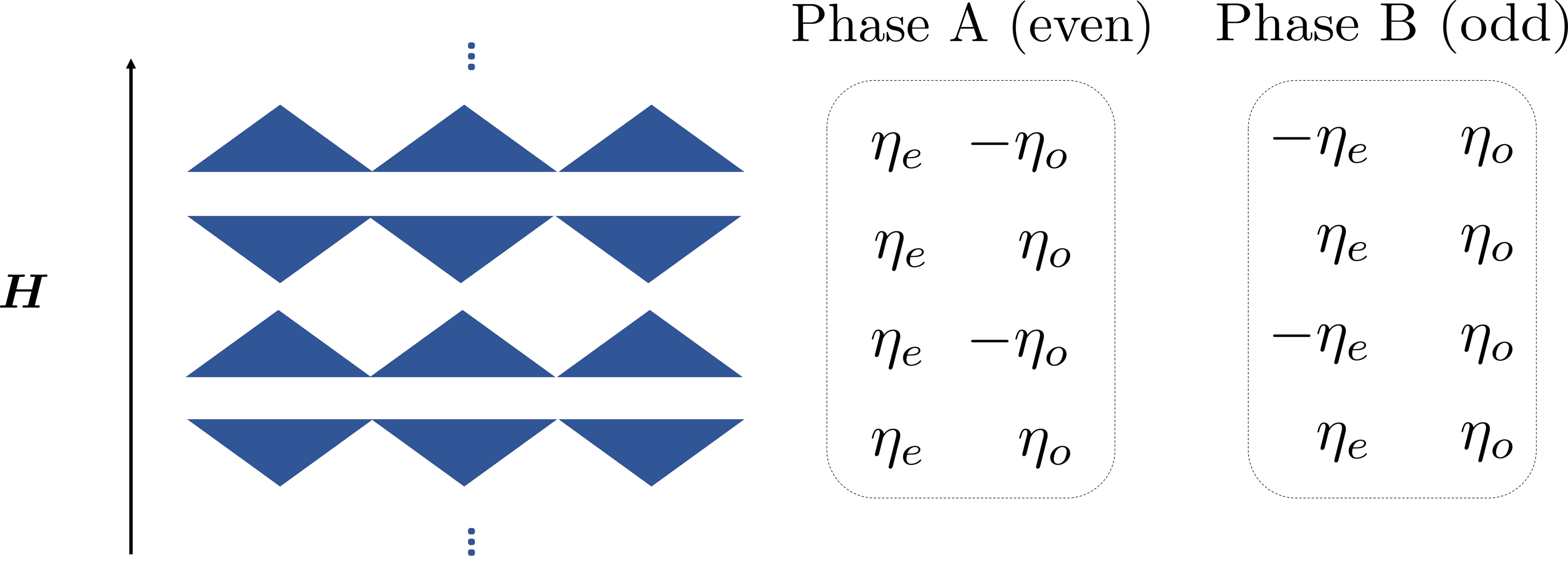}
\end{center}
\caption{Superconducting multilayer system motivated by CeRh$_2$As$_2$. The lack of local inversion symmetry, illustrated by the alternating orientation of the triangles, gives rise to Rashba effects causing the formation of Cooper pairs of mixed parity within each layer. The minimization of the interlayer coupling causes the dominant order parameter component to maintain its sign, whereas the other is forced to adopt the alternating sign of the spin-orbit coupling. The A and B order parameter structure shown are the solutions permitted by symmetry, where the latter phase is favored at high magnetic fields.} 
\label{fig: orderparameter}
\end{figure}

At the onset of superconductivity, both order parameters become non-zero, and the resulting state is invariant under time-reversal operation. Furthermore, the relative phase of the two order parameter components is gauge-invariant and anti-symmetric under time-reversal operation $ \mathcal{T} $, such that~\cite{JPSJ.83.044712}
\begin{equation}
    \left(\varphi_{e,j} - \varphi_{o,j}\right) \,\mathrm{mod} \,2\pi =  0, \, \pi.
    \label{eq: phasedifference}
\end{equation}
To minimizes the interlayer coupling, the dominant order-parameter component maintains its sign across all layers, whereas the subdominant alternates in sign on neighboring layers~\cite{JPSJ.81.034702}. Thus, the allowed order-parameter structures are either globally even under inversion, $\Psi_{e,j} = \Psi_e$, $\Psi_{o,j} = (-1)^j \Psi_o$, or globally odd, $\Psi_{e,j} = (-1)^j\Psi_e$, $\Psi_{o,j} = \Psi_o$ (see Fig.~\ref{fig: orderparameter}). The global parity is established by an inversion center lying between two layers and the corresponding parity operation exchanges two layers as well as changes the sign of the odd-parity order parameter component. 
To avoid confusion, we refer to the (globally) even order-parameter structure, which is favored at low fields, as phase A and the odd, high-field phase as phase B.

Finally, the phase difference $\varphi_{e,j}-\varphi_{o,j}$ adapts on each layer such that the overall free energy is minimal for the given order parameter structure, in particular, 

\begin{equation}
    f_{eo}^{(j)} = - \tilde{\epsilon} \psi_{e,j} \psi_{o,j}.
\end{equation}

In order to limit the number of free parameters, we assume several coefficient in the free energy to be the same for both order-parameter components, in particular, $b =b_e=b_o$, $\tilde{J} = \tilde{J}_e = \tilde{J}_o$ and $m = m_e = m_o$. Throughout this article, we adopt a dimensionless formulation of the free energy, unless stated otherwise. As detailed in Refs.~\onlinecite{PhysRevB.64.064517, PhysRevB.97.144508}, we measure lengths in units of the penetration depth $\lambda(T) = [m c^2 b /(16\pi e^2 a_e(T))]^{1/2}$ and re-scale the magnetic field by $\sqrt{2} B_c$, with $B_c$ the thermodynamic critical field. Because $B_c = \Phi_0/(2 \sqrt{2} \pi \lambda \xi)$, with the Ginzburg-Landau coherence length $\xi$ given as 
\begin{equation}
    \xi^2(T)= \frac{\hbar^2}{2m a_e(T)},
\end{equation}
this is equivalent to re-scaling the vector potential by $\Phi_0/[2\pi \xi(T)]$, where $\Phi_0 =  h c / 2e$ represents the flux quantum. In this formulation, the superconducting flux quantum is given by $\phi_0 =2\pi/\kappa_0$ and the dimensionless upper critical field $B_{c2}$ (orbital depairing field) is equal to $\kappa_0$, with $\kappa_0 = \lambda / \xi$ the temperature-independent Ginzburg-Landau parameter. Lastly, we normalize the order-parameter components by $\sqrt{a_e(T)/b}$ and denote them in dimensionless units by $\Xi_{l,j}(\bm{\rho},\varphi_{l,j}) = \eta_{l,j}(\bm{\rho})\exp(i\varphi_{l,j})$. 
In summary, we use
\begin{eqnarray}
  \bm{r} &\mapsto& \lambda(T) \bm{\rho},\\
  \bm{A}(\bm{r}) &\mapsto& \frac{\Phi_0}{2\pi \xi(T)} \mathcal{A}(\bm{\rho}),\\
  \Psi_{l,j}(\bm{r}) &\mapsto& \sqrt{a_e(T)/b} \Xi_{l,j}(\bm{\rho}).
\end{eqnarray}
The dimensionless free energy functional $\mathcal{F}$ is related to the free energy $F$ through
\begin{equation}
    F[\Psi_e, \Psi_o, \bm{A}] = \frac{a_e^2}{b} \mathcal{F}[\Xi_e, \Xi_o, \mathcal{A}].
\end{equation}

We rewrite the free energy density of Eq.~\eqref{eq: free_energy_density} in dimensionless form, and
 split it up into the following five parts:
\begin{equation}
    f^{(j)} = f_b^{(j)}+f_m^{(j)} + f_J^{(j)} + f_p^{(j)} + f_{eo}^{(j)},
\end{equation}
where the dimensionless basic free energy is given as
\begin{eqnarray}
f_b^{(j)}(\bm{\rho}) &=& -\eta_{e,j}^2(\bm{\rho}) -\frac{a_o}{a_e} \eta_{o,j}^2(\bm{\rho})+\frac{\eta_{e,j}^4(\bm{\rho})}{2} +\frac{\eta_{o,j}^4(\bm{\rho})}{2} \nonumber\\
&+& \frac{\left[\nabla \eta_{e,j}(\bm{\rho})\right]^2}{\kappa_0^2}+\frac{\left[\nabla \eta_{o,j}(\bm{\rho})\right]^2}{\kappa_0^2}.
\end{eqnarray}
The Ginzburg-Landau parameter $\kappa_{0}$ is equal for both even and odd components under the assumptions mentioned above. 
The term quadratic in the triplet component retains an explicit temperature dependence as the order parameter is normalized by the singlet quantity $a_e = a_{0,e} (T_{c,e}-T)$.

Next, we write the free energy density of the magnetic orbital part as
\begin{eqnarray}
f_m^{(j)}(\bm{\rho}) &=& \frac{\mathcal{B}^2(\bm{\rho})}{1+4\pi\chi_n}+ \left(\frac{\nabla \varphi_{e,j}}{\kappa_0}-\mathcal{A}(\bm{\rho})\right)^2 \eta_{e,j}^2(\bm{\rho})\nonumber\\
&+&\left(\frac{\nabla \varphi_{o,j}}{\kappa_0}-\mathcal{A}(\bm{\rho})\right)^2 \eta_{o,j}^2(\bm{\rho})\nonumber \\
&=&\mathcal{A}^2(\bm{\rho})\left[\eta_{e,j}^2(\bm{\rho}) + \eta_{o,j}^2(\bm{\rho})\right] + \frac{\mathcal{B}^2(\bm{\rho})}{1+4 \pi \chi_n},
\end{eqnarray}
where $\mathcal{B}(\bm{\rho})=\nabla \times \mathcal{A}(\bm{\rho})$ is the dimensionless magnetic induction and $\chi_n$ represents the normal-state susceptibility. Note that in the last line we exploited the constant phase difference, Eq.~\eqref{eq: phasedifference}, leading to $\nabla\varphi_{e,j} = \nabla \varphi_{o,j}$ and performed the gauge transformation $\mathcal{A} \rightarrow \mathcal{A} + \nabla \varphi/\kappa_0$~\cite{PhysRevB.97.144508}.
 
Finally, the interlayer coupling energy $f_J^{(j)}$, the paramagnetic contribution $f_p^{(j)}$ and the parity mixing $f_{eo}^{(j)}$ can be written as
\begin{eqnarray}
    f_J^{(j)}(\bm{\rho}) &=& J\big(|\eta_{e,j+1}(\bm{\rho})-\eta_{e,j}(\bm{\rho})|^2\nonumber\\
    &+& |\eta_{o,j+1}(\bm{\rho})-\eta_{o,j}(\bm{\rho})|^2 \big), \\
    f_p^{(j)}(\bm{\rho}) &=& \mathcal{Q} \chi_n \mathcal{B}^2 \eta_{e,j}^2(\bm{\rho}),\\
    f_{eo}^{(j)}(\bm{\rho}) &=& (-1)^{j+1} \epsilon \eta_{e,j}(\bm{\rho}) \eta_{o,j}(\bm{\rho}),
\end{eqnarray}
with $J$, $\epsilon$ and $\mathcal{Q}$ corresponding to the coupling constants $\tilde{J}$, $\tilde{\epsilon}$, $\tilde{Q}$ in the dimensionless formulation. The interlayer coupling parameter has the unit of an energy (per area) and is therefore re-scaled in the following way:
\begin{eqnarray}
   \frac{a_e^2}{b} f_J^{(j)}(\bm{\rho}) &=& \tilde{J}\sum_l |\psi_{l,j+1}(\bm{r})-\psi_{l,j}(\bm{r})|^2\nonumber \\
    &=& \frac{\tilde{J}}{\lambda^2} \frac{a_e}{b} \sum_l |\eta_{l,j+1}(\bm{\rho})-\eta_{l,j}(\bm{\rho})|^2 \nonumber\\
    &\eqqcolon& \frac{a_e^2}{b}  J\sum_l|\eta_{l,j+1}(\bm{\rho})-\eta_{l,j}(\bm{\rho})|^2 \nonumber \\
    %
\end{eqnarray} 
and thus $J= \tilde{J}/(a_e\lambda^2) = 2\tilde{J}/(\hbar \kappa_0)^2$. For the same reasoning, the spin-orbit coupling strength transforms as $\epsilon = 2 \tilde{\epsilon}/(\hbar \kappa_0)^2$. Lastly, the phenomenological constant $\mathcal{Q}$ is related to $\tilde{Q}$ through $\mathcal{Q}(T)=  \tilde{Q} a_e/b$, resulting in an explicit temperature dependence of $f_p^{(j)}$ in the dimensionless formulation. For simplicity we denote $\mathcal{Q}(0)=Q$.

In our formulation, the Maki parameter can be expressed through the system parameters, the normal state susceptibility $\chi_n$ and Ginzburg-Landau parameter 
\begin{equation}
    \alpha_M = \kappa_0 \sqrt{8 \pi \chi_n (1+4 \pi \chi_n) }.
\end{equation}
Imposing the order-parameter structure for the A and B phases, one can directly evaluate the sum over the layer index $j$. Note that
\begin{equation}
    f_J(\bm{\rho}) = 4 J
    \begin{cases}
    \eta_o^2(\bm{\rho}), \quad \mathrm{A \,phase}\\
    \eta_e^2(\bm{\rho}), \quad \mathrm{B \,phase}
    \end{cases}
\end{equation}
is the only term in the free energy that includes the order parameter phase difference between the layers.

In the remainder of the paper we choose the parameter values $\kappa_0 = 100$, $\alpha_M = 20$, $\epsilon = J =1$, $a_{0,e}=1$, $a_{0,o}=0.4$, $b=1$, $T_{c,o}/T_{c,e}=0.6$, and $\mathcal{Q} =0.1 \cdot 4 \pi a_e/b $ unless otherwise stated.

\subsection{Superconducting instability - linearized GL equations}

Before accounting for the spatial modulations of the order parameter due to the flux-line lattice, we solve the linearized GL equations to obtain the onset of superconductivity in a magnetic field. For phase A, these equations read
\begin{eqnarray}
0&=&\left[ \mathcal{Q} \chi_n \mathcal{B}^2 -1 + \left(-\frac{i}{ \kappa_0} \nabla + \mathcal{A}\right)^2\right] \Xi_e - \frac{\epsilon}{2} \Xi_o ,\\
0&=&\left[4J-\frac{a_o}{a_s} + \left(-\frac{i}{ \kappa_0} \nabla + \mathcal{A}\right)^2 \right] \Xi_o -\epsilon  \Xi_e ,
\end{eqnarray}
and for phase B
\begin{eqnarray}
0&=&\left[ \mathcal{Q} \chi_n \mathcal{B}^2 -1+4J + \left(-\frac{i\nabla}{ \kappa_0}  + \mathcal{A} \right)^2 \right] \Xi_e - \epsilon \Xi_o ,\\
0&=&\left[-\frac{a_o}{a_e} + \left(-\frac{i}{ \kappa_0} \nabla + \mathcal{A} \right)^2 \right]\Xi_o -\epsilon  \Xi_e. 
\end{eqnarray}

\begin{figure}[h]
\begin{center}
\includegraphics[width=1 \linewidth]{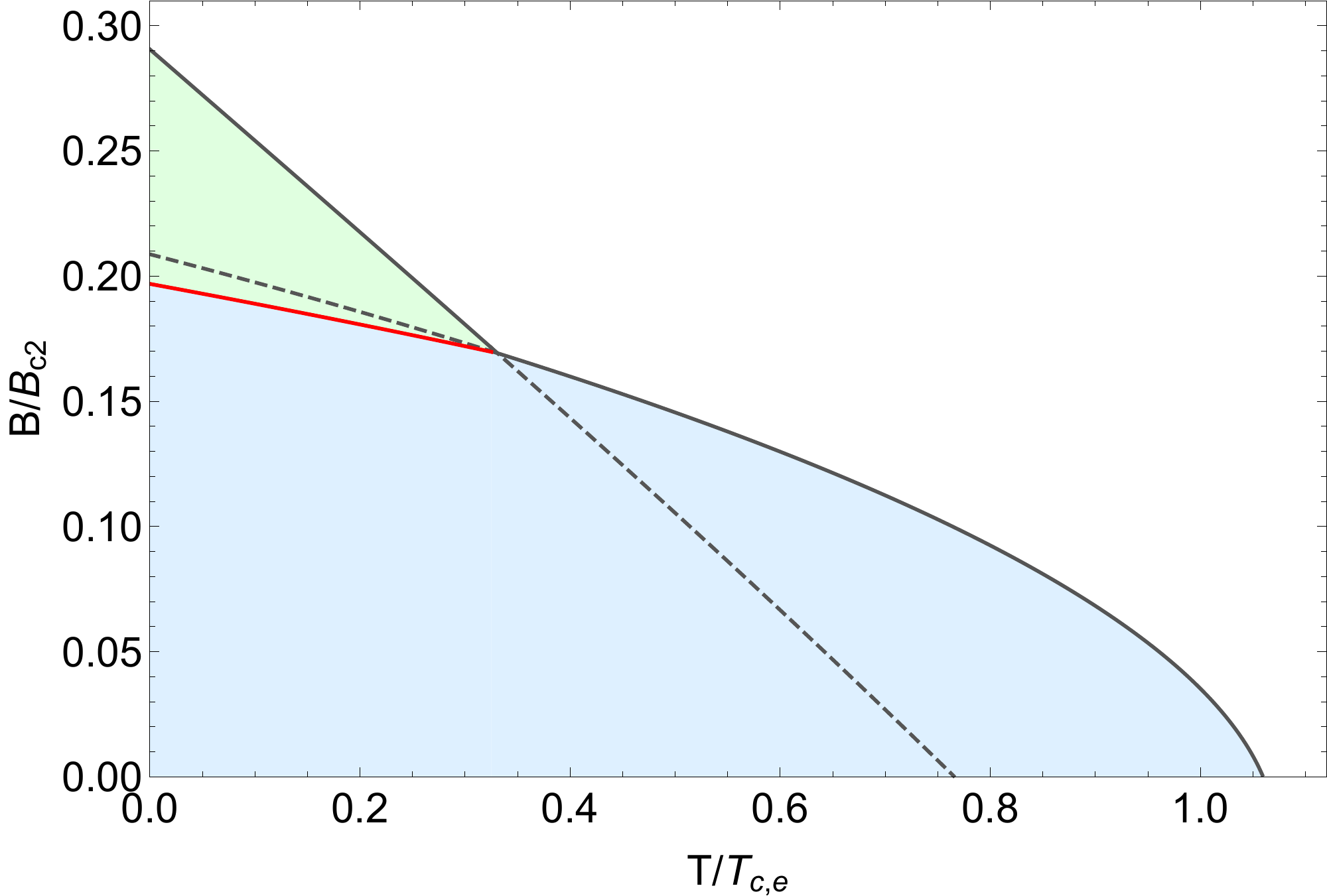}
\end{center}
\caption{$B$-$T$ phase diagram obtained from the linearized coupled GL equations. On the vertical axis, $B$ is scaled with $B_{c2}$ and the temperature axis is measured in units of the bare critical temperature of the even component $T_{c,e}$. The outer phase boundaries drawn in solid gray are in excellent agreement with the critical field found for the full numerical treatment in Fig.~\ref{fig: Hc_vs_T_multilayer}. The dashed lines mark the continuations of the crossing phase boundary lines with the A (blue) and B (green) phase and do not represent physical phase transitions. The red line corresponds to the critical field for the internal phase transition between A and B phase as derived in Sec.~\ref{subsec: System subject to paramagnetic limiting}.} 
\label{fig: lin_GL}
\end{figure}

For $ \bm{\mathcal{B}} = (0,0,\mathcal{B}) $, we choose the gauge $\mathcal{A} = ( 0, \mathcal{B}x,0)$ and use the ansatz $\Xi = \mathrm{e}^{i k_y y} u_\Xi(x)$, such that we can solve
\begin{equation}
\left(-i \kappa_0^{-1} \nabla + \mathcal{A}\right)^2 \Xi_e \eqqcolon E \Xi_e,
\end{equation}
giving
\begin{equation}
-u''(x) + (\mathcal{B}\kappa_0 x - k_y)^2 = \kappa_0^2 E \Xi_e,
\end{equation}
and leading to the Landau levels $E_n = \mathcal{B} (2n + 1)/\kappa_0$. The lowest Landau level is $E_0 = \mathcal{B}/\kappa_0$, and thus, the critical field for the A phase is determined by solving the equation,
\begin{eqnarray}
\mathrm{det}\begin{pmatrix}
\frac{\mathcal{B}}{\kappa_0} + \mathcal{Q}\chi_n \mathcal{B}^2 -1 & -\epsilon/2\\
-\epsilon/2 & \frac{\mathcal{B}}{\kappa_0} + 4J -\frac{a_o}{a_e}
\end{pmatrix}=0.
\end{eqnarray}
An analogous equation is obtained for the B phase. The resulting $B$-$T$ phase diagram is shown in Fig.~\ref{fig: lin_GL} and agrees well with the upper critical field of the model developed in Sec.~\ref{sec: Full treatment}.

\subsection{Internal phase transition}
\label{subsec: System subject to paramagnetic limiting}

As paramagnetic limiting only directly affects the singlet component, the low-field phase A is paramagnetically limited. On the other hand, the high-field phase B is dominated by the triplet component, which is relatively immune to paramagnetic pair breaking. As we consider systems with a large Maki parameter, we expect that orbital depairing only plays a subordinate role when discussing the transition from phase A to B. 

\begin{figure}[h]
\begin{center}
\includegraphics[width=1 \linewidth]{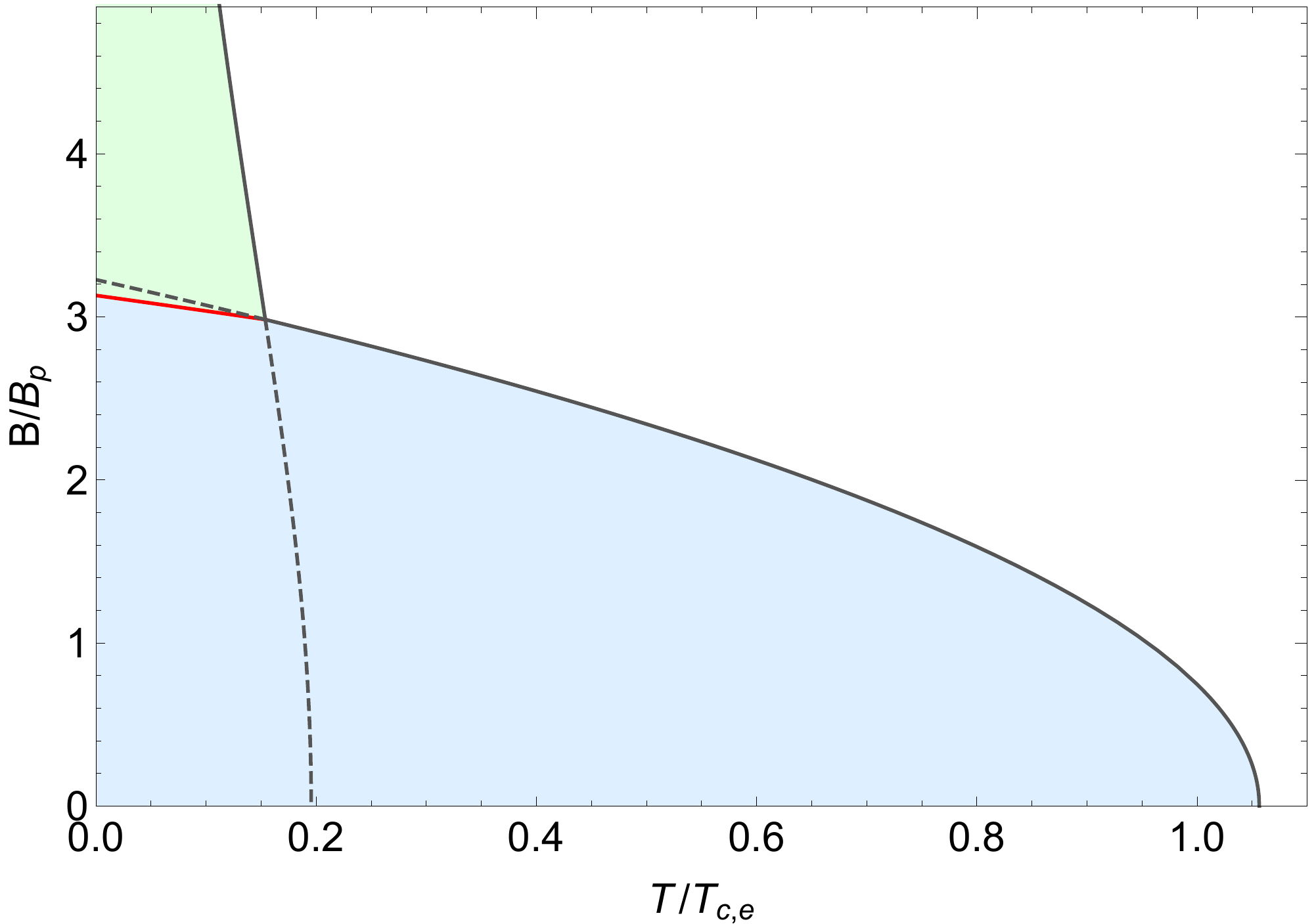}
\end{center}
\caption{$B$-$T$ phase diagram for the free energy including paramagnetic depairing as given in Eq.~\eqref{eq: F_toy_model_evaluated}. The field is measured in units of the zero-temperature paramagnetic critical field for a monolayer $B_p = (1+1/4\pi \chi_n)^{-1/2}$~\cite{PhysRevB.97.144508}. The A and B phases are shaded in blue and green, respectively. Note the dashed segment of the critical field of the A phase does not coincide with the critical field of the A-B transition (red line) described by Eq.~\eqref{eq: Bc_AB_transition}. For this figure, we chose $T_{c,o}=0$ and $Q \chi_n = 2 \cdot 10^{-3}$ to guarantee real solutions in the whole temperature range and have an effective Maki parameter $\alpha-{M} \approx 20$. 
} 
\label{fig: Hc_vs_T_toymodel}
\end{figure}

To gain an understanding of this internal phase transition, we thus use a simplification of Eq.~\eqref{eq: free_energy_density} that ignores the in-plane spatial modulations of the order parameter as well as the orbital depairing. For this analysis the terms second order in the order parameters are sufficient. The free energy for both phases is then written as\
\begin{eqnarray}
\label{eq: F_toy_model_evaluated}
F &= (-a_e + Q \chi_n \bm{B}^2) \psi_e^2 - a_o \psi_o^2 -\epsilon \psi_e\psi_o \nonumber\\
& \quad + 4 J \begin{cases}
\psi_o^2, \quad \mathrm{phase \, A}\\
\psi_e^2, \quad \mathrm{phase \, B}
\end{cases}.
\end{eqnarray}
Expressing the quadratic form through a symmetric matrix and imposing its determinant to be zero, we obtain the critical field for both phases
\begin{eqnarray}
\label{eq: toy_model_Bc_A}
\bm{B}_{c,\,\mathrm{A}}^2 &=& \frac{\epsilon^2 +4a_e\left(4J-a_o\right)}{4Q \chi_n (4J-a_o)},\\
\label{eq: toy_model_Bc_B}
\bm{B}_{c, \, \mathrm{B}}^2 &=& -\frac{\epsilon^2  +4a_o(4J - a_e)}{4Q \chi_n a_o},
\end{eqnarray}
the larger of which determines the field where superconductivity disappears (see Fig.~\ref{fig: Hc_vs_T_toymodel}). 

To determine the internal phase transition from phase A to phase B, their corresponding free energies are minimized and compared. As shown in App.~\ref{app_sec: Internal phase transition}, this leads to the expression for the critical field of the internal phase transition $\bm{B}_{c,\mathrm{AB}}$
\begin{equation}
    Q \chi_n \bm{B}_{c,\mathrm{AB}}^2 = a_e - a_o ,
    \label{eq: Bc_AB_transition}
\end{equation}
which highlights that the paramagnetic limiting effect is decisive in determining the internal transition by suppressing the dominant even-parity low-field in favor of a dominant odd-parity high-field pairing. 
Despite the significant simplifications, Eq.~\eqref{eq: Bc_AB_transition} is in good agreement with the phase diagram, including the effects of the vortex lattice in the mixed phase, computed in Sec.~\ref{sec: Full treatment}. 

Using Eq.~\eqref{eq: toy_model_Bc_A}, we also obtain the critical temperature at zero field
\begin{eqnarray}
T_{B_{c,\,\mathrm{A}}=0}&=& \frac{\sqrt{(4 J + a_{0,o}(T_{c,e}-T_{c,o}))^2 + \frac{a_{0,o}}{a_{0,e}} \epsilon^2}}{2 a_o}\nonumber\\
&+& \frac{T_{c,e}+T_{c,o}}{2}- \frac{2J}{a_{0,o}},
\label{eq: toy_model_Tc_B=0}
\end{eqnarray}
which for finite $\epsilon$ is larger than $T_{c,e}$ and converges to $T_{c,e}$ for $\epsilon\rightarrow0$. Similarly, by equating~\eqref{eq: toy_model_Bc_A} with~\eqref{eq: toy_model_Bc_B} we obtain the temperature at which the two critical fields meet,

\begin{eqnarray}
T_{B_{c,\,\mathrm{A}}=B_{c,\,\mathrm{B}}} &=& T_{c,o} + \frac{\sqrt{16J^2+ \epsilon^2}-4J}{2 a_{0,o}} \label{eq: toy_model_Tc_A=B_approx}\\
&\approx& T_{c,o} + \frac{\epsilon^2}{16 a_{0,o} J}.\nonumber
\end{eqnarray}
 Equation~\eqref{eq: toy_model_Tc_A=B_approx} indicates that the spin-orbit coupling also increases the temperature range in which the B phase can be observed. Furthermore, it shows the importance of the interplay between spin-orbit coupling and the interlayer coupling for the formation of the high-field B phase.


\section{Vortex lattice in the mixed phase}
\label{sec: Full treatment}
\subsection{Circular cell method}

To analyze the situation for a regular flux-line lattice, we employ the vortex cell method introduced by Brandt \textit{et al.}~\cite{PhysRevB.64.064517} and previously used for a trilayer system~\cite{PhysRevB.97.144508}. 
For this purpose, we decompose the system in plane into a triangular lattice of hexagonal cells each containing one vortex in the center. Following the Wigner-Seitz idea, we approximate the cells by a circle of the same area $ A_{\rm{cell}} $ as the hexagon. The free energy restricted to a single cell with the corresponding boundary conditions then reads
\begin{equation}
F[\Psi_e,\Psi_o, \bm{A}] = \frac{1}{N A_{\rm{cell}}} \sum_j \int_{\rm{cell}} d^2r \; f^{(j)} (\bm{r}) 
\end{equation}
with $ N $ the number of layers. This expression can be converted into a dimensionless formulation as done before.

The cell approach is based on the trial function for the vortex structure~\cite{clem1975simple},
\begin{equation}
\eta_l^2 (\rho) = \eta_{l,\infty} ^2 \frac{\rho^2}{\rho^2 + \xi_c^2}, 
\label{eq: var_ansatz}
\end{equation}
where $l=e,o$ and $\eta_{e,\infty}$, $\eta_{o,\infty}$, and $\xi_c$ are treated as variational parameters for the even and odd bulk magnitude as well as the vortex core size. 

The trial functions, Eq.~\eqref{eq: var_ansatz}, reduce the system to a single Ginzburg-Landau equation, obtained by the variational derivative with respect to $\mathcal{A}$, which reads
\begin{equation}
\frac{\nabla^2 \mathcal{A}}{1+ 4\pi\chi_n} = \Delta_\infty^2 \frac{\rho^2}{\rho^2 + \xi_\Delta ^2} \mathcal{A},
\label{eq: free_energy_var_wrt_A}
\end{equation}
where we defined
\begin{eqnarray}
\Delta_\infty ^2 &=& \frac{\eta_{e,\infty}^2 + \eta_{o,\infty}^2}{1+\chi_n (1+4\pi\chi_n) Q \eta_{e,\infty}^2}, \\
\xi_\Delta ^2 &=& \frac{\xi_c^2}{1+\chi_n (1+4\pi\chi_n) Q \eta_{e,\infty}^2}.
\end{eqnarray}
By taking the curl of Eq.~\eqref{eq: free_energy_var_wrt_A}, one obtains the differential equation~\cite{saint-james1969}
\begin{equation}
\frac{1}{1+4\pi\chi_n}\frac{1}{\rho}\dfrac{d}{d\rho} \left[\frac{1}{\Delta_\infty^2} \frac{\rho^2+\xi_\Delta^2}{\rho} \dfrac{d\mathcal{B}}{d\rho} \right] = \mathcal{B},
\label{eq: free_energy_diff_eq_for_B}
\end{equation}
whose general solution is given in Ref.~\cite{PhysRevB.64.064517},
\begin{eqnarray}
\mathcal{B} =\frac{ \sqrt{\eta_{e,\infty}^2 + \eta_{o,\infty}^2}}{\kappa_0 \xi_c } \frac{c_1 K_0\left( \Delta_\infty P\right) + c_2 I_0\left( \Delta_\infty P \right)}{1+4\pi\chi_n} .
\end{eqnarray}
Here, $I_n$ and $K_n$ are the $n$th-order modified Bessel functions of first and second kind, respectively, and we define $P^2 = \rho_\mathcal{B}^2 + \xi_\Delta^2$ with $\rho_\mathcal{B}$ being the radius of the vortex cell, i.e., $A_\mathrm{cell} = \pi \rho_\mathcal{B}^2 = 2\pi/(\kappa_0 \bar{\mathcal{B}})$ with $\bar{\mathcal{B}}$ the mean magnetic induction in a vortex core. The coefficients $c_1$ and $c_2$ are determined through the boundary conditions of vanishing current density at the cell boundary as well  as the flux threading the unit cell equalling a flux quantum $\phi_0$ (for $\chi_n = 0$). The latter is formalized by integrating the left side of Eq.~\eqref{eq: free_energy_diff_eq_for_B},
\begin{equation}
\phi_0 = \frac{2\pi}{\kappa_0} = 2 \pi \int_0^{\rho_\mathcal{B}} \mathrm{d}\rho\, \rho \mathcal{B} (\rho),
\end{equation}
which together with the first boundary condition leads to
\begin{equation}
-\frac{1}{\kappa_0} = \lim_{\rho\rightarrow 0} \left[\frac{1}{\Delta_\infty^2} \frac{\rho^2+\xi_\Delta^2}{\rho} \dfrac{d\mathcal{B}}{d\rho}\right].
\end{equation}
Using $\chi_n \ll 1$, the coefficients $c_1$ and $c_2$ reduce to
\begin{align}
c_1  &=  \frac{I_1(\Delta_\infty P)}{I_1(\Delta_\infty P) K_1(\Delta_\infty \xi_\Delta )- I_1(\Delta_\infty \xi_\Delta) K_1(\Delta_\infty P)},\\
c_2 &=  \frac{K_1(\Delta_\infty P)}{I_1(\Delta_\infty P) K_1(\Delta_\infty \xi_\Delta )- I_1(\Delta_\infty \xi_\Delta ) K_1(\Delta_\infty P)}.
\end{align}
With this analytic expression for the magnetic induction, we can use the result by Hao and Clem for the magnetic part, $\mathcal{F}_m = \bar{\mathcal{B}}\mathcal{B}(0)$~\cite{PhysRevB.43.2844}. 

Next, we integrate the free energy density within the cell which leads to the analytical expressions for the different parts of the free energy,
\begin{eqnarray}
\mathcal{F}_b &=& -\mathcal{C}_k(\bar{\mathcal{B}}, \xi_c) \left(\eta_{e,\infty}^2 + \frac{a_o}{a_e} \eta_{o,\infty}^2 \right) \nonumber \\
&+& \frac{\bar{\mathcal{B}}(1+\bar{\mathcal{B}}\kappa_0 \xi_c^2)}{\kappa_0(2+\bar{\mathcal{B}} \kappa_0 \xi_c^2)^2} (\eta_{e,\infty}^2 + \eta_{o,\infty}^2) \nonumber   \\
&+&\left[\mathcal{C}_k(\bar{\mathcal{B}}, \xi_c) -\frac{1}{2+\bar{\mathcal{B}} \kappa_0 \xi_c^2} \right] (\eta_{e,\infty}^4+\eta_{o,\infty}^4),\\
\mathcal{F}_m &=& \frac{I_1(\Delta_\infty P)K_0(\Delta_\infty \xi_\Delta)+ I_0(\Delta_\infty \xi_\Delta)K_1(\Delta_\infty P)}{I_1(\Delta_\infty P) K_1(\Delta_\infty \xi_\Delta )- I_1(\Delta_\infty \xi_\Delta ) K_1(\Delta_\infty P)}\nonumber \\
&\times&\frac{\bar{\mathcal{B}}}{1+4\pi\chi_n} \frac{\sqrt{\eta_{e,\infty}^2 + \eta_{o,\infty}^2}}{\kappa_0 \xi_c}, \\
\mathcal{F}_J &=& 4  \mathcal{C}_k(\bar{\mathcal{B}}, \xi_c) J \begin{cases}
\eta_{o,\infty}^2, \quad\mathrm{phase \, A},\\
\eta_{e,\infty}^2,\quad \mathrm{phase\, B},
\end{cases}\\
\mathcal{F}_p &=& \mathcal{C}_k(\bar{\mathcal{B}}, \xi_c)  Q \chi_n \bar{\mathcal{B}}^2 \eta_{e,\infty}^2, \\
\label{eq: free_energy_paramagnetic}
\mathcal{F}_{eo} &=& -\mathcal{C}_k(\bar{\mathcal{B}}, \xi_c) \epsilon \eta_{e,\infty} \eta_{o,\infty},
\end{eqnarray}
and we introduced
\begin{equation}
    \mathcal{C}_k(\bar{\mathcal{B}}, \xi_c) = 1+ \frac{\bar{\mathcal{B}} \kappa_0 \xi_c^2}{2} \ln \left(1- \frac{2}{2+\bar{\mathcal{B}} \kappa_0 \xi_c^2} \right).
\end{equation}
In the following, we minimize the the free energy numerically using the Nelder-Mead algorithm with respect to the variational parameters $\eta_{e,\infty}$, $\eta_{o,\infty}$, and $\xi_c$.

\begin{figure}[t]
\begin{center}
\includegraphics[width= \linewidth]{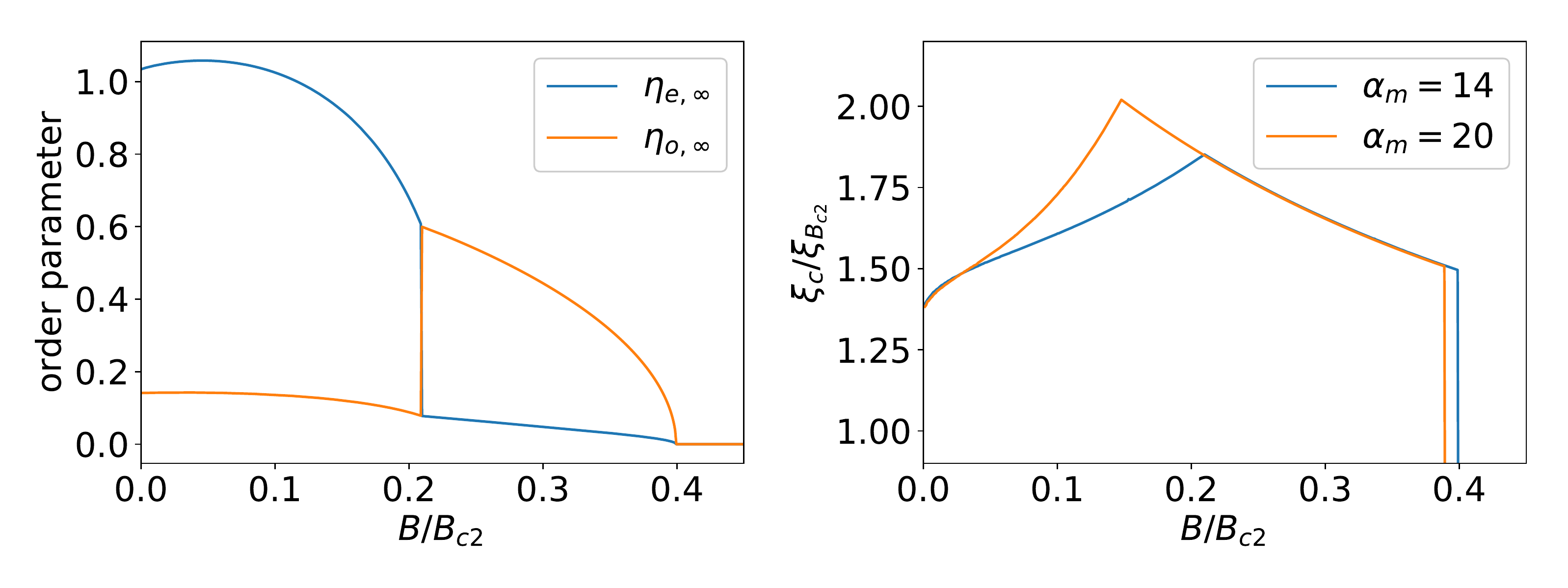}
\end{center}
\caption{Numerical minimization of the variational parameters $\eta_{e,\infty}$, $\eta_{o,\infty}$ (left) and $\xi_c$ (right). The magnetic induction  is normalized by the orbital upper orbital critical field $B_{c2} = \kappa_0$ and the vortex core size by $\xi_{B_{c2}} = 1/\kappa_0$, its value at $B_{c2}$. The parameters $\eta_{e,\infty}$ and $\eta_{o,\infty}$ are shown for a Maki parameter $\alpha_M=14$, while $\xi_c$ is additionally shown for $\alpha_M =20$. The initial increase in $\eta_{e,\infty}$ is the result of a numerical artefact.} 
\label{fig: variational_parameters_vs_B_1J}
\end{figure}

\subsection{Results}
\label{subsec: Results}

We choose a large Ginzburg-Landau parameter of $\kappa_0 = 100$, ensuring that the lower critical field $\mathcal{H}_{c1}$ is very small compared to the upper critical field. The numerical results for the order-parameter magnitudes, as well as the vortex-core size as a function of the field (at zero temperature), are displayed in Fig.~\ref{fig: variational_parameters_vs_B_1J}. The numerical results show that at low fields, the singlet-dominated phase A is indeed favored. This changes at sufficiently high fields, where the odd component becomes competitive.

\begin{figure}[b]
\begin{center}
\includegraphics[width=\linewidth]{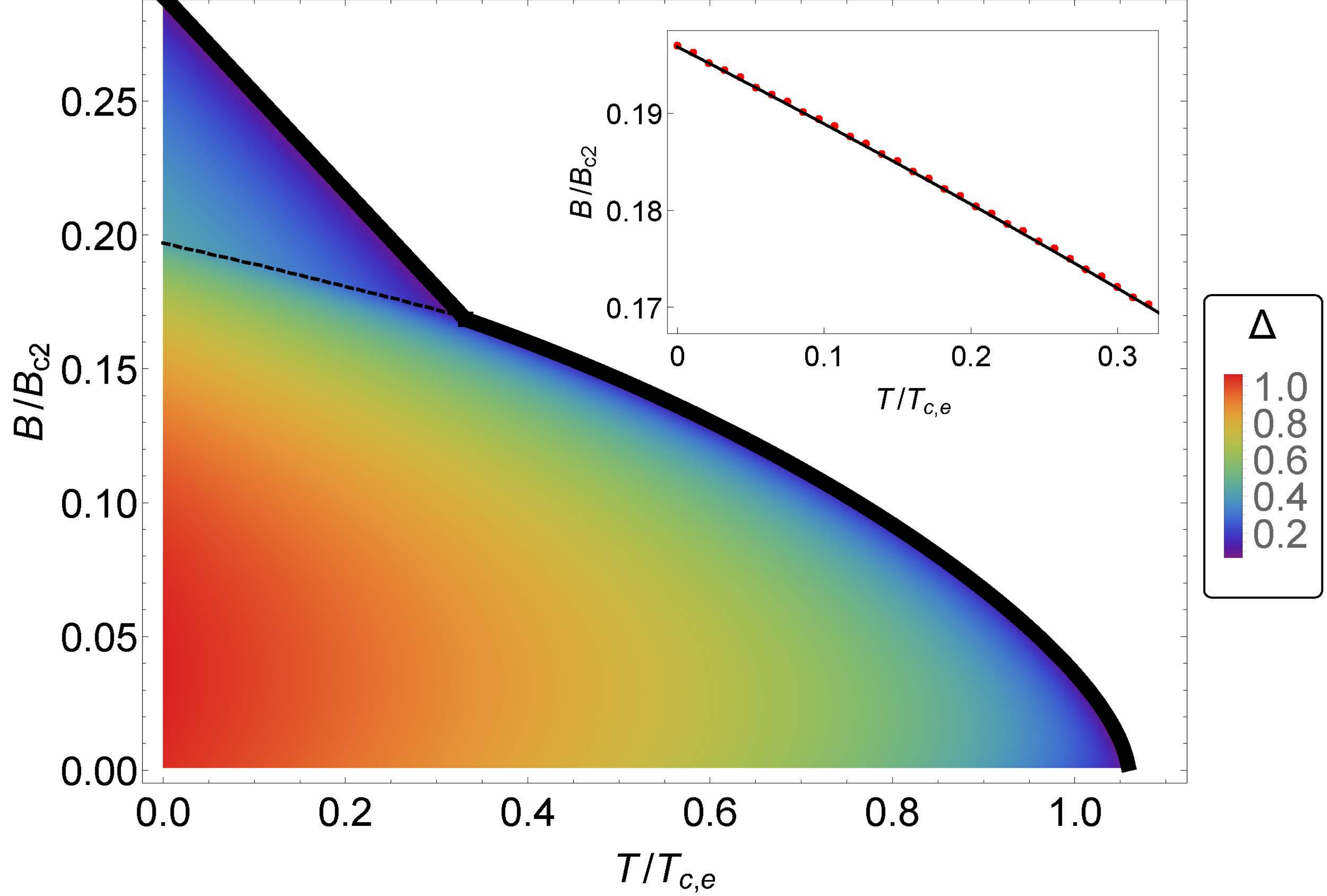}
\end{center}
\caption{$B$-$T$ phase diagram with the solid black line showing the second-order phase transition. The color map displays the amplitude of the superconducting order parameter, $\Delta = \sqrt{\eta_{e,\infty}^2 + \eta_{o,\infty}^2}$, and the dashed black line marks the A-B transition. The inset shows the comparison of the numerically determined critical field for the A-B transition (red points) with the analytical approximation in Eq.~\eqref{eq: Bc_AB_transition}. } 
\label{fig: Hc_vs_T_multilayer}
\end{figure}

Figure~\ref{fig: Hc_vs_T_multilayer} shows a typical $B$-$T$ phase diagram, obtained by reinstating the explicit temperature dependence. The two outer phase boundaries are of second order, whereas the internal phase boundary is of first order. The inset shows the comparison of this critical field $B_{c, AB}$ with the one in Eq. ~\eqref{eq: Bc_AB_transition} derived from the simplified free energy. The excellent agreement suggests that orbital depairing and the flux line lattice only play a minor role in determining the critical field for the internal phase transition, which is dominated by the paramagnetic limiting effect. 

\begin{figure}[ht]
\begin{center}
\includegraphics[width= \linewidth]{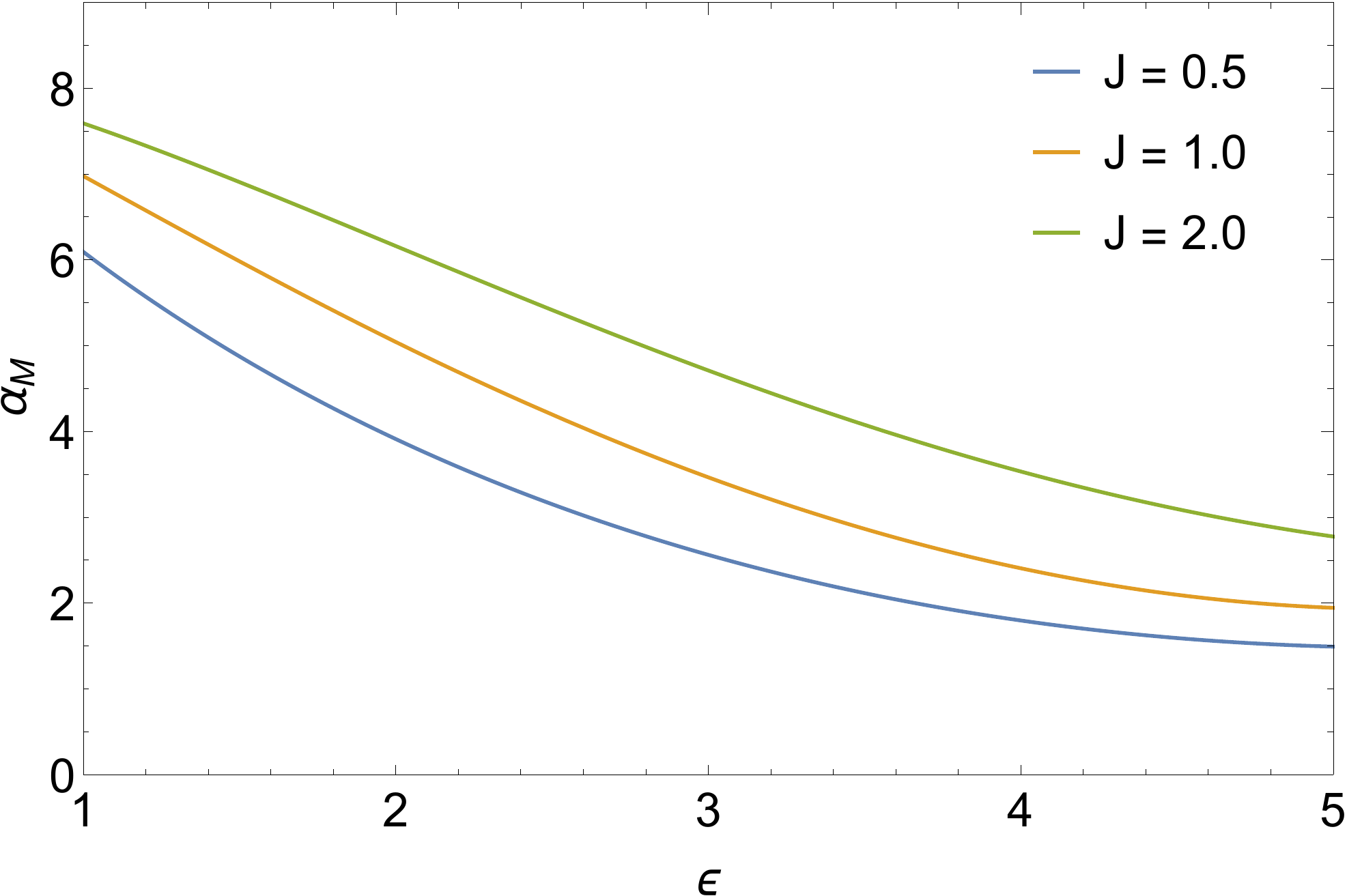} 
\end{center}
\caption{Minimal Maki parameter $\alpha_M$ required for phase B to form at $T=0$ as a function of the spin-orbit coupling strength $\epsilon$ for different interlayer coupling strengths $J$. A strong spin-orbit coupling is beneficial for phase B to form, whereas the interlayer coupling stabilizes phase A. Here we used $a_{0,o}=0.6$ and $\mathcal{Q} = 0.15 \cdot 4 \pi a_e/b$.} 
\label{fig: Maki_vs_epsilon}
\end{figure}

Although the expression of the critical field for the internal phase transition, Eq.~\eqref{eq: Bc_AB_transition}, does not explicitly contain the spin-orbit or Josephson interlayer coupling constants $\epsilon$ and $J$, they play a crucial role in the formation of this high-field phase. This is illustrated in Fig.~\ref{fig: Maki_vs_epsilon}, where the minimal Maki parameter required for phase B to form decreases with increasing spin-orbit coupling strength. The opposite applies to $J$, as a strong interlayer coupling suppresses the admixed subdominant odd order parameter component in phase A, thereby diminishing the effect of the spin-orbit-induced parity mixing.

\begin{figure}
\begin{center}
\includegraphics[width= \linewidth]{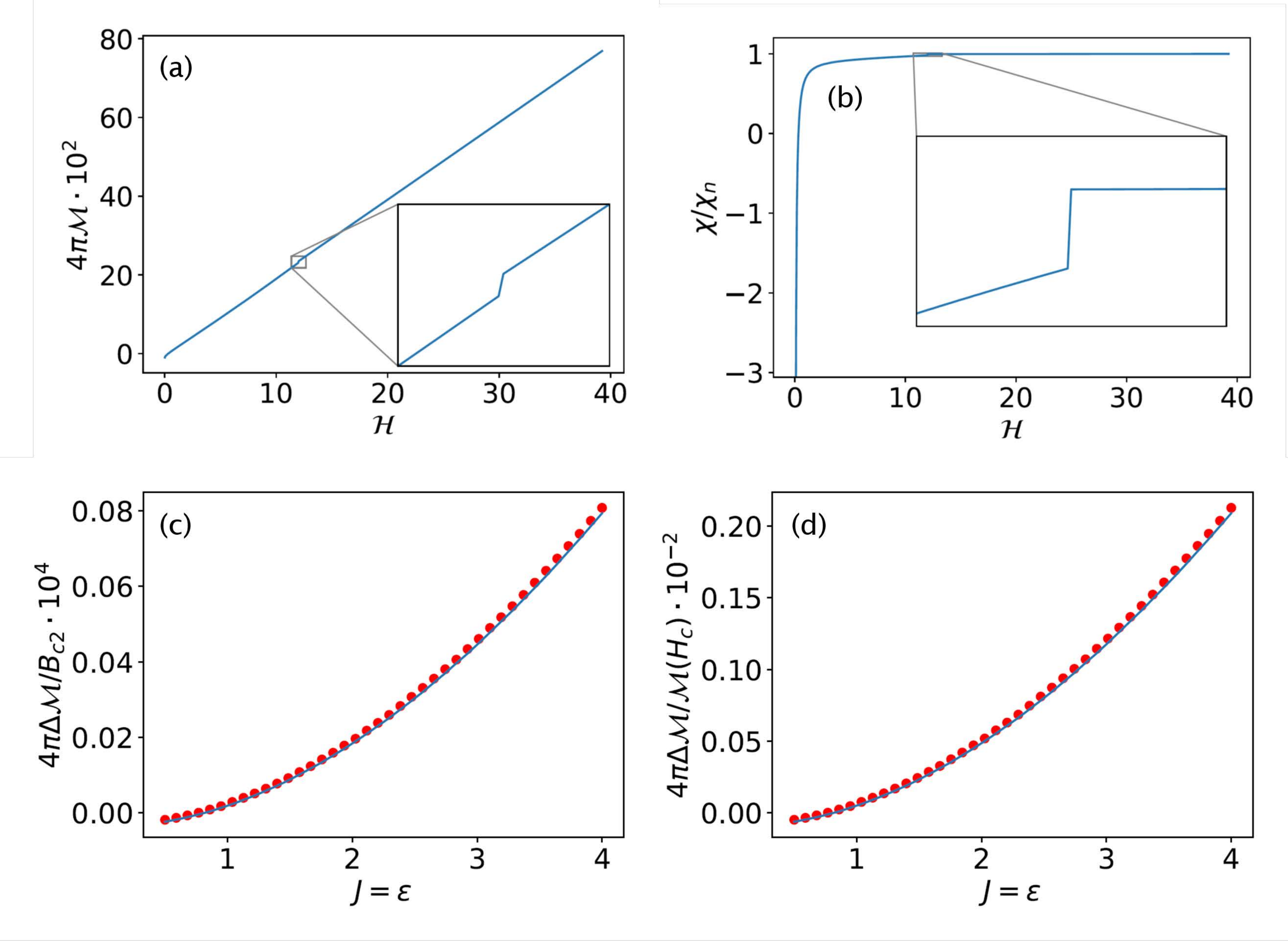} 
\caption{(a) Typical magnetization and (b) susceptibility curves. Both curves show a jump at the internal first-order phase transition. The relative jump in magnetization is on the order of $1\%$ for typical parameter values. (c) Absolute  and (d) relative magnetization jump $\Delta \mathcal{M}$ as a function of Josephson coupling $J$, which in this plot is set equal to the spin-orbit coupling strength. The blue lines correspond to the approximation in Eq.~\eqref{eq: magnetization_jump_s}, whereas the red points are results obtained from the numerical solution of the full model. The magnetization jump increases with growing $J=\epsilon$. For the plots (a) and (b) we used $a_{o,0}=0.8$ and $T_{c,o}/T_{c,e} = 0.7$, whereas we used the usual parameter values for plots (c) and (d).}
\label{fig: magnetization_and_chi}
\end{center}
\end{figure}

A possible way to experimentally observe the A-B transition is by measuring the magnetization curve. To this end, we calculate the dimensionless magnetic field $\mathcal{H}$ using
\begin{equation}
\mathcal{H} = \frac{1}{2} \dfrac{\partial \mathcal{F}}{\partial \bar{\mathcal{B}}}.
\end{equation}
The analytic expression for $\mathcal{H}$ is included in App.~\ref{app_sec: Magnetization}. The magnetization and magnetic susceptibility can then be extracted from $4 \pi \mathcal{M} = \mathcal{B}-\mathcal{H}$ and $\chi = \mathcal{M}/\mathcal{H}$. A typical magnetization and susceptibility curve is shown in Fig.~\ref{fig: magnetization_and_chi} showing a jump at the A-B transition and thus providing an experimental signature of the internal first-order phase transition. 

We can approximately express the magnetization jump by (see App.~\ref{eq_app: H_field})
\begin{equation}
\Delta \mathcal{M} = \sqrt{Q\chi_n\left(1-\frac{a_o}{a_e}\right)}  \left(\eta_{e,\infty}^2-\eta_{o,\infty}^2\right) \left[\mathcal{C}_k(\bar{\mathcal{B}}_c, \xi_c) -1/2\right].
\label{eq: magnetization_jump_s}
\end{equation}
Unlike the critical field for the internal phase transition, Eq.~\eqref{eq: magnetization_jump_s} explicitly contains the variational parameters $\psi_\infty$, $\eta_\infty$, and $\xi_c$ and can, thus, not be directly evaluated. We observe that the magnetization jump increases with growing $ J $ and $ \epsilon$. Consequently, increasing the interlayer coupling strength $J$ and/or the spin-orbit coupling $\epsilon$ leads to a larger magnetization jump as the relative importance of the orbital depairing term shrinks. This behavior can be seen in Fig.~\ref{fig: magnetization_and_chi}, where we compared the analytical approximation of Eq.~(\ref{eq: magnetization_jump_s}) with the full numerical solution.

\begin{figure}
\begin{center}
\includegraphics[width= \linewidth]{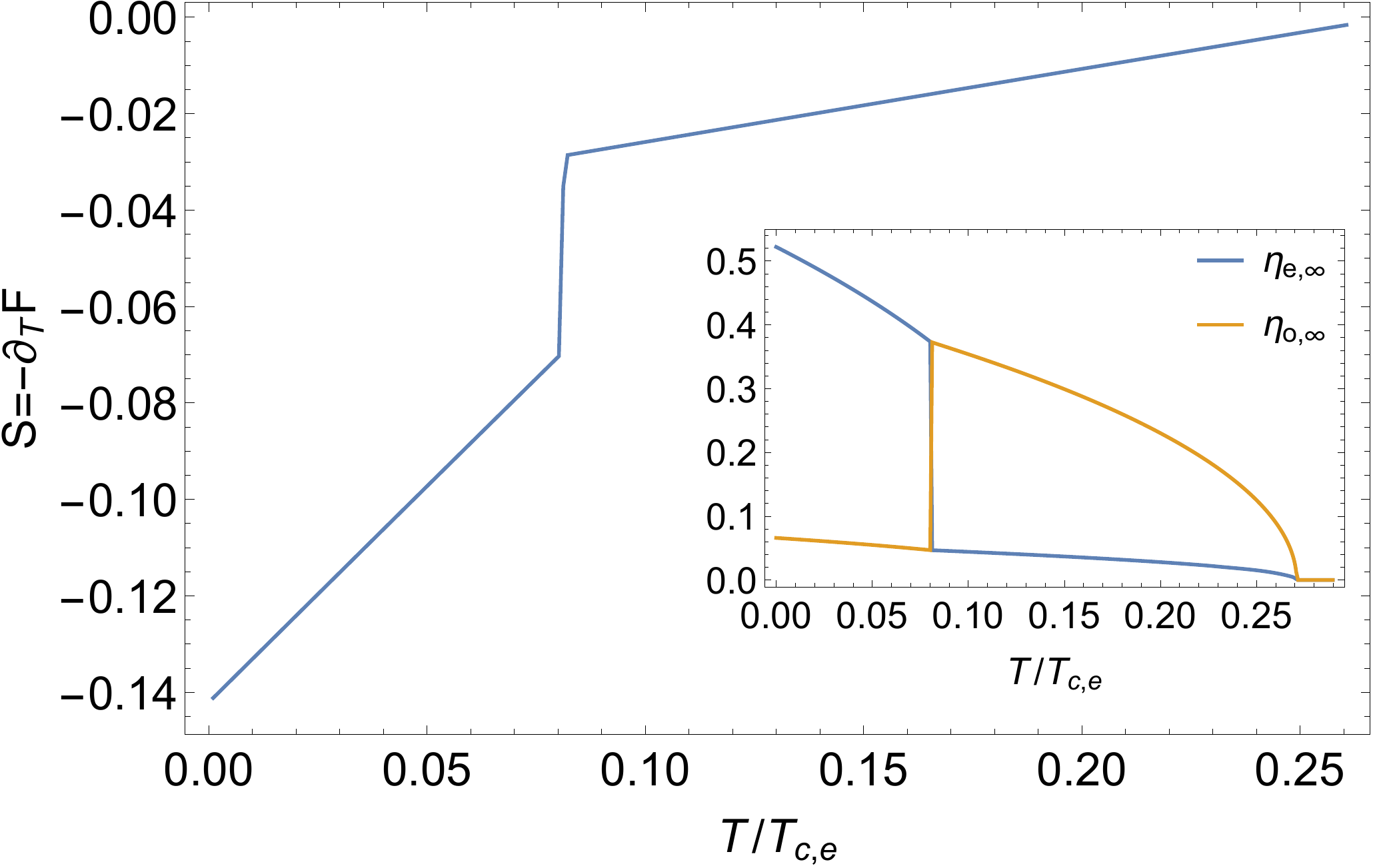} 
\caption{Entropy as a function of temperature at $B/B_{c2}=0.187$. The entropy shows a discontinuity in accordance with the first-order nature of the internal transition. 
 Inset: order parameters $\eta_{e,\infty}$ and $\eta_{o,\infty}$ as a function of temperature at constant field showcasing the swapping of order parameters at $T_c$.}
\label{fig: entropy_vs_T_single_J}
\end{center}
\end{figure}

Apart from the phase diagram and magnetization curve, a further experimentally accessible quantity is the latent heat. However, it must be noted that observing the A-B transition by changing the temperature at constant magnetic field may prove difficult for typical parameter values, as the critical field $\bar{\mathcal{B}}_c$ only shows a weak temperature dependence.
To obtain the thermodynamic quantities, we must reintroduce the explicit temperature dependence. As expected for a first-order transition, the entropy $S = -\partial_T F$ is discontinuous at the internal phase transition, see Fig.~\ref{fig: entropy_vs_T_single_J}. 

\section{Discussion}

A previous study by Yoshida et al. has investigated non-centrosymmetric bi- and trilayer systems within the framework of Bogoliubov-de-Gennes~\cite{PhysRevB.86.134514}. They similarly find that at low fields, the sign of the singlet order parameter remains constant due to interlayer coupling, whereas at high fields, a phase forms in which the order parameter adapts to the sign of the staggered Rashba spin-orbit coupling. Furthermore, while the results for the trilayer system show certain additional peculiar features due to the centrosymmetric middle layer, the resulting phase diagram for the bilayer system is qualitatively similar to Fig.~\ref{fig: Hc_vs_T_multilayer}. However, the state they investigated neglected the in-plane spatial modulations of the order parameter in the mixed phase. Thus, important aspects of the internal phase transition, including the magnetization jump and the role of the Maki parameter, could not be thoroughly investigated.

The two anomalies at the internal phase boundary line, the jump in the magnetization and the entropy (latent heat), are not the only features which may be used to observe this phase transition. Additionally, we expect that anomalies are observable in the ac-susceptibility as well as in ultrasound velocity (a change in the elastic constants) for certain modes. A further intriguing feature could result from the fact that in the B phase, the odd-parity exceeds the even-parity component. It has been proposed that in this case helical subgap edge states would appear at the surfaces with in-plane normal vectors~\cite{PhysRevB.76.012501,PhysRevB.79.060505}. Such helical states would be visible in in-plane quasiparticle tunneling spectroscopy as zero-bias anomalies. In the A phase with dominant even-parity component, such helical edge states would be absent. Thus, we expect a clear qualitative change of the quasiparticle $I$-$V$-characteristics when passing through the transition. Finally, the experimental analysis of the magnetocaloric effect may allow to detect the internal phase transition~\cite{PhysicaB.163.155}.


\section{Conclusion}

Staggered non-centrosymmetric superconductors with Rashba spin-orbit coupling alternating in sign from layer to layer involve an order parameter of mixed parity which can be influenced by external magnetic fields. While an even-parity spin-singlet pairing state may dominate over an odd-parity spin-triplet state at low magnetic fields, paramagnetic limiting effects may lead to a suppression of the former in favor of the latter in a growing magnetic field. This may trigger a phase transition to a phase with dominant odd-parity component provided that the Maki parameter is sufficiently large, i.e. that orbital depairing is weak due to a very short coherence length. 

We suggest that such a situation is realized in the recently found heavy-Fermion superconductor CeRh$_2$As$_2$, which shows an anomaly in the upper critical field which exceeds the expected paramagnetic limiting field drastically. We show that besides the kink-like anomaly of the upper critical fields to the normal state, there is also an internal phase transition within the mixed phase between the low-field phase A with dominant even-parity and the high-field phase B with dominant odd-parity pairing. This internal phase boundary may be observed by various means as detailed in the previous section. In particular, this phase boundary should connect to the kink of the upper critical field. An intriguing feature of the B phase is the fact that it may have subgap edge states observable in quasi-particle tunneling spectroscopy. 

We have not considered here features expected for inplane magnetic fields. As discussed in Ref.~\cite{JPSJ.82.074714}, there may exist a high-field superconducting phase, which has a spatially modulated order parameter similar to the helical phase~\cite{PhysRevLett.92.027003, JETP.78.401, PhysRevLett.94.137002}. 
Also in this case, the comparative strength of spin-orbit and interlayer coupling is essential, while the Maki parameter does not play an essential role.

\begin{acknowledgments}
We would like to thank Daniel F.~Agterberg, Elena Hassinger, Seunghyun Khim, Andrew P.~Mackenzie, David M\"ockli, and Youichi Yanase for many enlightening discussions. This work was financially supported by a Grant of the Swiss National Science Foundation (No.184739). 
\end{acknowledgments}

\vspace{10pt}


\appendix


\section{Internal phase transition}
\label{app_sec: Internal phase transition}

At the internal phase transition in the $B$-$T$ phase diagram, the free energy in Eq.~\eqref{eq: F_toy_model_evaluated} is equal for both phases and minimal with respect to the corresponding order parameters. In this section we denote the order parameter values which minimizes the free energy of phase A, $F_A$, as $\psi_{A,e}$, $\psi_{A,o}$, and analogously use $\psi_{B,e}$, $\psi_{B,o}$ for phase B. First, we calculate

\begin{eqnarray}
0&=&\psi_{B,e} \dfrac{\partial F_A}{\partial \psi_{A,e}} - \psi_{A,e} \dfrac{\partial F_B}{\partial \psi_{B,e}} \nonumber \\
&=&-\epsilon \left(\psi_{A,o} \psi_{B,e}-\psi_{A,e}\psi_{B,o}\right) -8 J \psi_{A,e} \psi_{B,e}, \label{eq: app_Internal phase transition1} \\
0&=&\psi_{B,o} \dfrac{\partial F_A}{\partial \psi_{A,o}}-\psi_{A,o} \dfrac{\partial F_B}{\partial \psi_{B,o}}  \nonumber \\
&=&-\epsilon \left( \psi_{A,e} \psi_{B,o}-\psi_{A,o} \psi_{B,e} \right) + 8 J \psi_{A,o}\psi_{B,o}. \label{eq: app_Internal phase transition2}
\end{eqnarray}
By adding Eq.~\eqref{eq: app_Internal phase transition1} and~\eqref{eq: app_Internal phase transition2} one obtains
\begin{equation}
    \psi_{A,e}\psi_{B,e} = \psi_{A,o} \psi_{B,o}.\label{eq: app_Internal phase transition3}
\end{equation}
Using the extremal condition of $F_A$ with respect to $\psi_{A,e}$ and of $F_B$ with respect to $\psi_{B,o}$ we obtain
\begin{equation}
    \frac{-a_e + Q \chi_n \bm{B}^2}{-a_o} \frac{\psi_{A,e}}{\psi_{B,o}} = \frac{\psi_{A,o}}{\psi_{B,e}},
\end{equation}
which together with Eq.~\eqref{eq: app_Internal phase transition3} yields the expression for the critical field in Eq.~\eqref{eq: Bc_AB_transition}
\begin{equation}
    Q \chi_n \bm{B}^2 = a_e -a_o.
\end{equation}

Finally, we note that for equal interlayer coupling for the even- and odd-parity 
order parameter, they simply swap value at the internal phase transition. This can be obtained by equating both the interlayer and spin-orbit free energy contributions for the two phases, underlining the key roles these two terms play in the A-B transition. 


\begin{widetext}
\section{Magnetization}
\label{app_sec: Magnetization}

For better readability, we omitted the expression for the $\bm{H}$-field in the main text. The full analytic expression reads

\begin{eqnarray}
2\mathcal{H} &=& -\dfrac{\partial \mathcal{C}_k}{\partial \bar{\mathcal{B}}} (\eta_{e,\infty}^2 + \frac{a_o}{a_e} \eta_{o,\infty}^2)  + \frac{2+ 3 \bar{\mathcal{B}} \kappa_0 \xi_c^2}{\kappa_0 (2+ \bar{\mathcal{B}} \kappa_0 \xi_c^2)^3} (\eta_{e,\infty}^2 + \eta_{o,\infty}^2)+ \frac{(K_1(\Delta_\infty \xi_\Delta) I_1(\Delta_\infty P) -K_1(\Delta_\infty P) I_1 (\Delta_\infty \xi_\Delta))^{-2}}{(1+4\pi\chi_n) \bar{\mathcal{B}} \kappa_0^2 \xi_c^2 P^2}  \nonumber  \nonumber \\
&+& \frac{\sqrt{\eta_{e,\infty}^2+\eta_{o,\infty}^2}}{\kappa_0 \xi_c(1+4\pi\chi_n)} \frac{K_0(\Delta_\infty \xi_\Delta) I_1 (\Delta_\infty P)+ I_0(\Delta_\infty \xi_\Delta) K_1 (\Delta_\infty P)}{K_1(\Delta_\infty \xi_\Delta) I_1(\Delta_\infty P)- I_1(\Delta_\infty \xi_\Delta) K_1 (\Delta_\infty P)} + \frac{\kappa_0 \xi_c^2}{2} \left[ \frac{2(3+ \bar{\mathcal{B}} \kappa_0 \xi_c^2)}{(2+ \bar{\mathcal{B}}\kappa_0 \xi_c^2)^2}+\ln\left(1-\frac{2}{2+\bar{\mathcal{B}}\kappa_0 \xi_c^2}\right)\right]  \nonumber\\
&\cdot&(\eta_{e,\infty}^4+\eta_{o,\infty}^4)+  Q \chi_n \bar{\mathcal{B}}\eta_{e,\infty}^2 \left( 2\mathcal{C}_k + \bar{\mathcal{B}} \dfrac{\partial\mathcal{C}_k}{\partial \bar{\mathcal{B}}}\right)+\dfrac{\partial \mathcal{C}_k}{\partial \bar{\mathcal{B}}} \epsilon \eta_{e,\infty} \eta_{o,\infty}+ 4 J \dfrac{\partial \mathcal{C}_k}{\partial \bar{\mathcal{B}}}  \begin{cases}
\eta_{o,\infty}^2, \, \, \mathrm{phase \, A}\\
\eta_{e,\infty}^2, \, \, \mathrm{phase \, B}
\end{cases},
\label{eq_app: H_field}
\end{eqnarray}
where the derivative of $\mathcal{C}_k$ is given as
\begin{equation}
\dfrac{\partial \mathcal{C}_k}{\partial \bar{\mathcal{B}}} = \frac{1}{2} \kappa_0 \xi_c^2 \left[\frac{2}{2+ \bar{\mathcal{B}} \kappa_0 \xi_c ^2}+ \ln\left(1-\frac{2}{2+\bar{\mathcal{B}}\kappa_0 \xi_c^2} \right) \right].
\end{equation}

We can calculate the jump in magnetization at the internal transition where $\bar{\mathcal{B}} = \bar{\mathcal{B}}_c$ in the case $J_e = J_o \eqqcolon J$. Then, as shown in App.~\ref{app_sec: Internal phase transition} the order parameters swap values at the first-order transition. Thus, only the terms originating from the magnetic orbital part, $\Delta_\infty$, $\xi_\Delta$, and $P$, change under the permutation $\eta_{e,\infty} \leftrightarrow \eta_{o,\infty}$. Therefore, it is useful to split magnetization jump in to two parts
\begin{equation}
\Delta \mathcal{M} = \Delta \mathcal{M}_m + \Delta \mathcal{M}_0,
\end{equation}
where $\mathcal{M}_m$ is the magnetization jump arising from the orbital term and $\Delta \mathcal{M}_0$ the remaining parts. The latter can be easily calculated by exploiting the fact that the variational parameters $\eta_{e,\infty}$ and $\eta_{o,\infty}$ swap values at the A-B transition

\begin{eqnarray}
2\Delta \mathcal{H}_1 = -\dfrac{\partial \mathcal{C}_k}{\partial \bar{\mathcal{B}}}\Bigr\rvert_{\bar{\mathcal{B}} = \mathcal{B}_c} \left(\eta_{o,\infty}^2+\frac{a_o}{a_e} \eta_{e,\infty}^2-\eta_{e,\infty}^2 - \frac{a_o}{a_e} \eta_{o,\infty}^2\right)+ \mathcal{B}_c Q \chi_n \bigg(2\mathcal{C}_k+\mathcal{B}_c \dfrac{\partial\mathcal{C}_k}{\partial\bar{\mathcal{B}}}\bigg\rvert_{\bar{\mathcal{B}} = \mathcal{B}_c}\bigg)  (\eta_{o,\infty}^2-\eta_{e,\infty}^2)
\end{eqnarray}

Using the critical field in Eq.~\eqref{eq: Bc_AB_transition} we obtain the contribution to the magnetization jump
\begin{equation}
\Delta \mathcal{M}_0 = \mathcal{C}_k(\bar{\mathcal{B}}_c, \xi_c) \sqrt{Q\chi_n\left(1-\frac{a_o}{a_e}\right)} (\eta_{e,\infty}^2-\eta_{o,\infty}^2) >0.
\end{equation}

For the orbital contribution $\Delta \mathcal{M}_m$ we note that the term 
\begin{equation}
\frac{(K_1(\Delta_\infty \xi_\Delta) I_1(\Delta_\infty P) -K_1(\Delta_\infty P) I_1 (\Delta_\infty \xi_\Delta))^{-2}}{(1+4\pi\chi_n) \bar{\mathcal{B}} \kappa_0^2 \xi_c^2 P^2}
\end{equation}
has a strong dependence on the arguments of the modified Bessel functions as $K_1(x)$ diverges like $1/x$ near zero. The other originating from the magnetic orbital part, written on the second line of Eq.~\eqref{eq_app: H_field}, is almost unaffected by the permutation of the order parameters and thus neglected. Upon inspecting the definitions of $P^2$, $\Delta_\infty^2$, and $\xi_\Delta^2$ we observe that, as $Q \chi_n$, $\xi_c^2 << 1$, the change in $\xi_\Delta \approx \xi_c$ is very small when $\eta_{e,\infty}$ and $\eta_{o,\infty}$ swap values. Therefore, we ignore the effect of the permutation $\eta_{e,\infty} \leftrightarrow \eta_{o,\infty}$ on $P$ and $\xi_\Delta$. Furthermore, we only account for the change in $\Delta_\infty$ in the most singular term $K_1(x)$. 
Replacing the modified Bessel function of the second kind with its asymptotic behaviour near zero, $K_1(x) \approx 1/x$, the contribution to the magnetization jump can be written as

\begin{eqnarray}
\Delta\mathcal{M}_m \propto \frac{1}{ \bar{\mathcal{B}} \kappa_0^2 \xi_c^2 P^2} \Bigg[\bigg(\frac{1}{\Delta_\infty \xi_\Delta} I_1(\Delta_\infty P)-\frac{1}{\Delta_\infty P}I_1(\Delta_\infty \xi_\Delta)\bigg)^{-2}- \bigg(\frac{1}{\tilde{\Delta}_\infty \xi_\Delta} I_1(\Delta_\infty P)-\frac{1}{\tilde{\Delta}_\infty P}I_1(\Delta_\infty \xi_\Delta)\bigg)^{-2} \Bigg]
\label{app_eq: deltaM_m1}
\end{eqnarray}
where we omitted the pre-factor $1/2(1+4\pi\chi_n)$ for compactness and with $\tilde{\Delta}_\infty$ the same as $\Delta_\infty$ apart from $\psi_{e,\infty}$ and $\eta_{o,\infty}$ being permuted. Using $I_1(x) \approx x/2$ the term in the square bracket of Eq.~\eqref{app_eq: deltaM_m1} simplifies to

\begin{eqnarray}
  \Bigg[\bigg(\frac{P}{2\xi_\Delta} -\frac{ \xi_\Delta}{2  P}\bigg)^{-2}- \bigg(\frac{\Delta_\infty P}{2\tilde{\Delta}_\infty \xi_\Delta}-\frac{\Delta_\infty \xi_\Delta}{2\tilde{\Delta}_\infty P}\bigg)^{-2} \Bigg].
\end{eqnarray}
Reinserting the definitions of $P$, $\Delta_\infty$ and $\xi_\Delta$ and setting $\bar{\mathcal{B}}=\bar{\mathcal{B}}_c$ we obtain the total magnetization jump at the internal phase transition

\begin{eqnarray}
\label{eq: magnetization_jump}
\Delta \mathcal{M} = \sqrt{Q\chi_n \left(1-\frac{a_o}{a_e}\right)} \left(\eta_{e,\infty}^2-\eta_{o,\infty}^2\right) \bigg[\mathcal{C}_k(\bar{\mathcal{B}}_c, \xi_c)-\frac{1/2}{{(1+\chi_n (1+4\pi\chi_n) Q \eta_{e,\infty}^2)(1+\chi_n (1+4\pi\chi_n) Q \eta_{o,\infty}^2)}} \bigg],
\end{eqnarray}
which for $\chi_n<<1$ this expression is well approximated by 
\begin{equation}
\Delta \mathcal{M} = \sqrt{Q\chi_n \left(1-\frac{a_o}{a_e}\right) } (\eta_{e,\infty}^2-\eta_{o,\infty}^2) \left[\mathcal{C}_k(\bar{\mathcal{B}}_c, \xi_c) -\frac{1}{2}\right].
\label{eq: magnetization_jump_3}
\end{equation}
\end{widetext}

\bibliographystyle{apsrev4-2}
\bibliography{references}

\end{document}